\documentclass{PoS}
\usepackage{subfigure}
\title{GRB Cosmology and Self-organized Criticality in GRBs}

\ShortTitle{GRB Cosmology and SOC in GRBs}

\author{\speaker{F. Y. Wang}\thanks{A footnote may follow.}\\
        School of Astronomy and Space Science, Nanjing University\\
        E-mail: \email{fayinwang@nju.edu.cn}}

\abstract{Gamma-ray bursts (GRBs), which have isotropic energy up to
$10^{54}$ erg, would be the ideal tool to study the properties of
early universe: including dark energy, star formation rate, and the
metal enrichment history of the Universe. We will briefly review the
progress on the field of GRB cosmology. Meanwhile, X-ray flares,
which may have important clues to the central engine, are common
phenomena in the GRB afterglows. We present statistical results of
X-ray flares, i.e., energy, duration time and waiting time
distributions, and compare the results with solar flares. The
similarity between the two kinds of flares are found, which may
indicates that the physical mechanism of GRB X-ray flares is
magnetic reconnection.}

\FullConference{Swift: 10 Years of Discovery,\\
        2-5 December 2014\\
        La Sapienza University, Rome, Italy }

\begin{document}
\section{Introduction}
The accelerating Universe at the present epoch was discovered from
the observation of type Ia supernovae (SNe Ia)
\cite{Riess98,Perlmutter99}. The cosmic accelerated expansion has
also been confirmed by other observations, such as cosmic microwave
background (CMB) \cite{Komatsu11}, and large-scale structure (LSS)
\cite{Eisenstein05}. But the SNe Ia can only be detected at low
redshifts, i.e., $z<2.0$. Gamma-ray bursts (GRBs), as the most
powerful explosions in the Universe \cite{Zhang04,Wang15}, can fill
the gap between SNe and CMB. GRBs can serves as the complementary
tools to measure dark energy and cosmic expansion
\cite{Dai04,Ghirlanda04b,Liang05,Amati08,Wang11a,Wang11b}. Long GRBs
can be made ``relative standard candles", using luminosity
correlations that have been found in prompt and afterglow phases
\cite{Liang05,Amati02,Ghirlanda04a}.

X-ray flares with short rise and decay times are discovered by Swift
satellite. The isotropic-equivalent energy of X-ray flares is from
$\sim 10^{48}$ to $10^{52}$ erg. The occurrence times of X-ray
flares range from a few seconds to $10^6$ seconds after the GRB
trigger. Until now, the physical origin of X-ray flares has remained
mysterious, although some models have been proposed. Here we
investigate the energy release frequency distribution, duration-time
frequency distribution and waiting time distribution of GRB X-ray
flares \cite{Wang13}. On the other hand, it is well known that solar
flare is triggered by a magnetic reconnection process. So we will
compare these two kind of flares.

\section{Constraints on dark energy}
GRBs have been widely used to constrain dark energy.  We use the
calibrating method to standardize 151 long GRBs with $E_{\rm
iso}-E_{\rm p}$ correlation. Just as using Cepheid variables to
standardize SNe Ia, the GRBs can be calibrated with SNe Ia. We use
the latest Union 2.1 data. This method is also cosmological
model-independent \cite{Liang08}. The full GRB data is separated
into two groups. The dividing line is the highest redshift in SNe Ia
Union 2.1 data, namely $z=1.414$. The low-redshift group ($z<1.414$)
includes 61 GRBs, and the high-redshift group ($z>1.414$) contains
90 GRBs. The left panel of Figure 1 shows the 1$\sigma$ and
2$\sigma$ constraints on the $\Omega_m-\Omega_\Lambda$ plane. Adding
the GRB data can give much more tighter constraints.

We also use the nonparametric technique to constrain the
uncorrelated equation of state (EOS) of dark energy $w(z)$. The EOS
is the most important parameters of dark energy. We apply the
nonparametric method to a joint data set of the latest observations
including SNe Ia, CMB from Planck and WMAP polarization, large scale
structure, the Hubble parameter measurement and GRBs. The right
panel of Figure 1 shows the evolution of uncorrelated EOS of dark
energy, which is consistent with $\Lambda$CDM at the $2\sigma$
confidence level \cite{Wang14}.

\begin{figure}
\begin{center}
\includegraphics[width=0.45\textwidth]{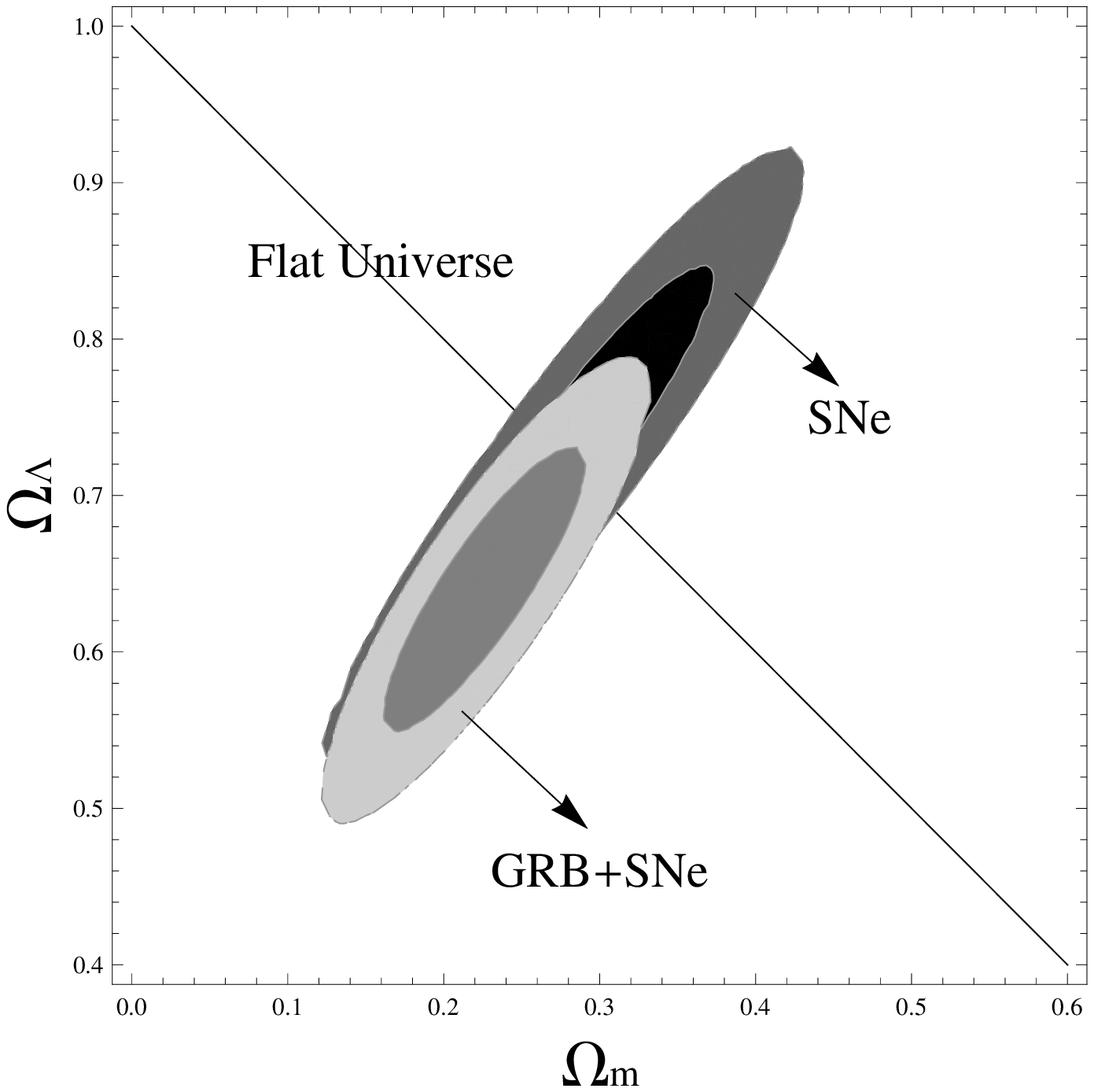}
\includegraphics[width=0.45\textwidth]{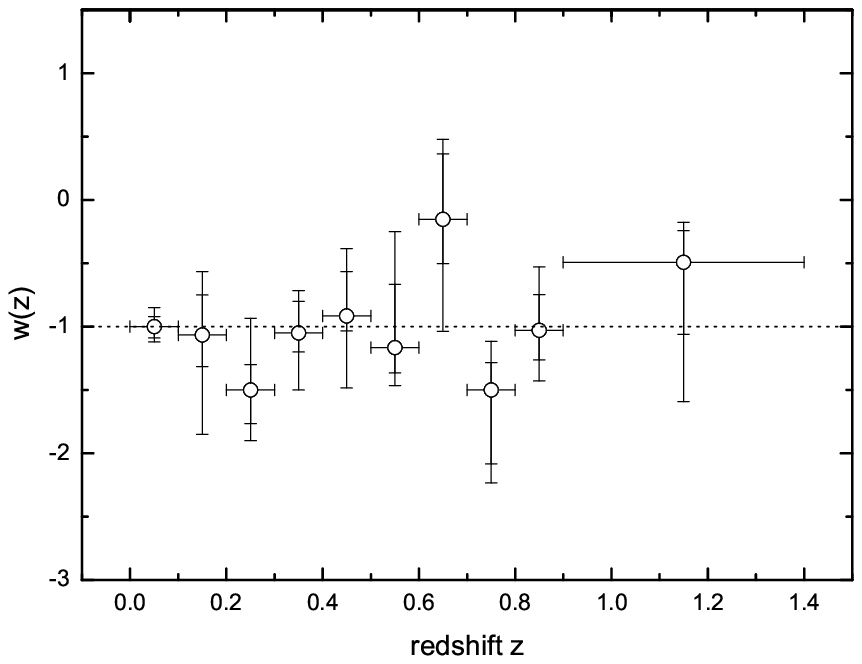}
\end{center}
\caption{Left: 1$\sigma$ and 2$\sigma$ constraints on $\Omega_m$ and
$\Omega_{\Lambda}$ from different data sets. The solid line shows
the $\Omega_k=0$ case. Right: Estimation of the uncorrelated dark
energy EOS parameters at different redshift bins
($w_1,w_2,...,w_{10}$) from SNe Ia+BAO+WMAP9+H(z) data. The open
points show the best fit value. The error bars are $1\sigma$ and
$2\sigma$ confidence levels. The dotted line shows the cosmological
constant. Adopted from \cite{Wang14}.}
\end{figure}

\section{Star formation rate}
\begin{figure}
\includegraphics[width=0.5\textwidth]{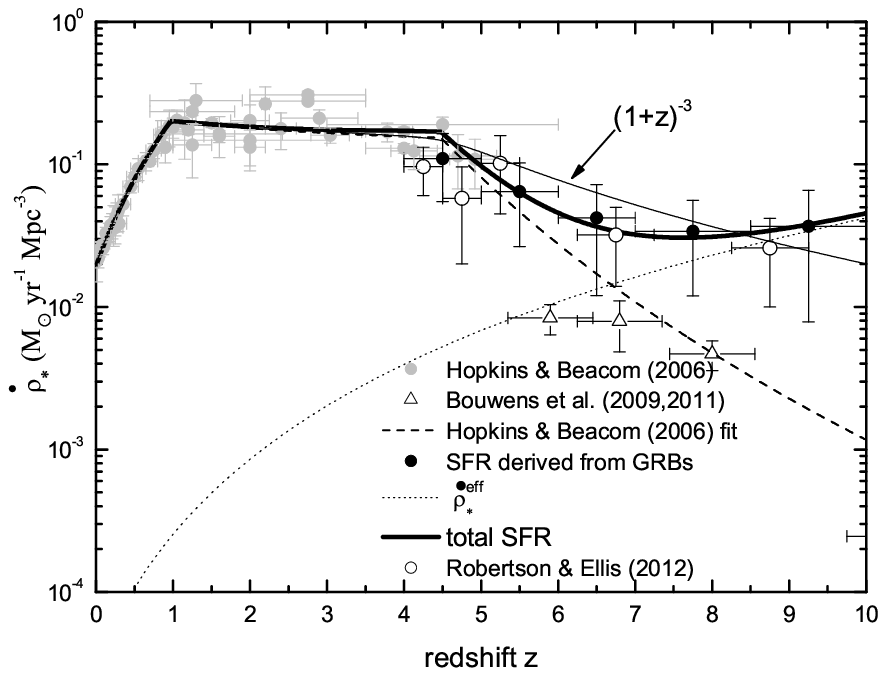}
\includegraphics[width=0.5\textwidth]{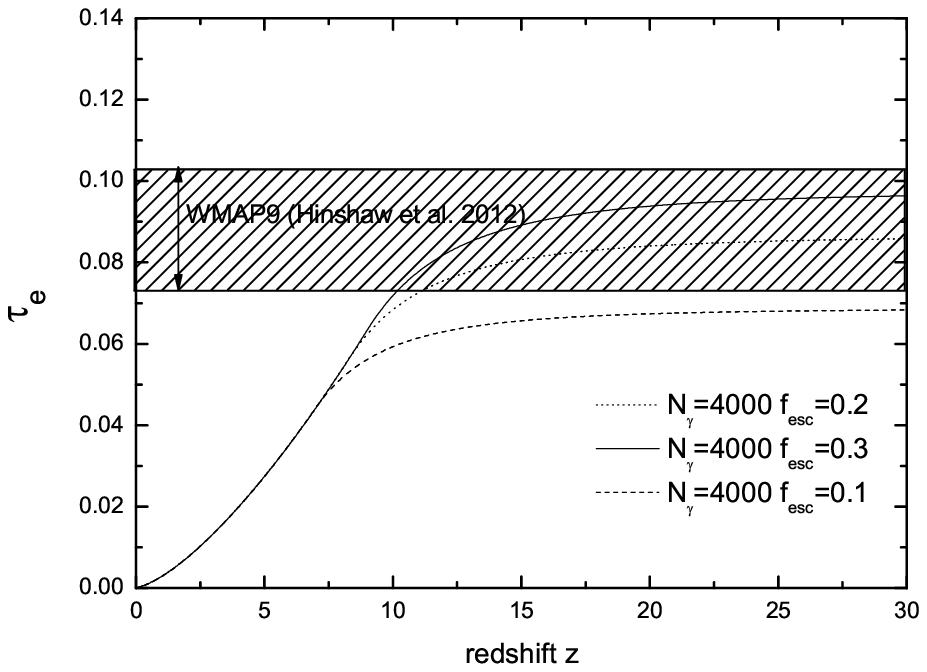}
\caption{The cosmic star formation history. The dashed line shows
their fitting result. The filled circles are the SFR derived from
GRBs in this work. Adopted from Wang (2014).} \label{SFR}
\end{figure}

The high-redshift star formation rate (SFR) is important in many
fields in astrophysics. Direct SFR measurement is still out of reach
by present instruments, particularly at the faint end of the galaxy
luminosity function. Long GRBs triggered by the death of massive
stars, provide a complementary technique for measuring the SFR. In
order to test the GRB rate relative to the SFR, we must choose
bursts with high luminosities, because only bright bursts can be
seen at low and high-redshifts, so we choose the luminosity cut
$L_{\rm iso}> 10^{51}$ erg s$^{-1}$ \cite{Yuksel08} in the redshift
bin $0-4$. This removes many low-redshift, low-$L_{\rm iso}$ bursts
that could not have been seen at higher redshift. The SFR can be
estimated as
\begin{equation}
\left\langle \dot{\rho}_* \right\rangle_{z_1-z_2} =
\frac{N_{z_1-z_2}^{\rm obs}}{N_{1-4}^{\rm obs}} \frac{\int_{1}^{4}
dz\, \frac{dV_{\rm com}/dz}{1+z}(1+z)^\delta \dot{\rho}_*(z)\,
}{\int_{z_1}^{z_2} dz\, \frac{dV_{\rm com}/dz}{1+z}(1+z)^\delta },
\label{zratio}
\end{equation}
where $N_{z_1-z_2}^{\rm obs}$ is the observed GRB counts, and
$\delta=0.5$ \cite{Wangf13}. The derived SFR from GRBs are shown as
filled circles in the left panel of Figure~\ref{SFR}. Error bars
correspond to 68\% Poisson confidence intervals for the binned
events. The high-redshift SFRs obviously decrease with increasing
redshifts, although an oscillation may exist. We find that the SFR
at $z>4.48$ is proportional to $(1+z)^{-3}$ using minimum $\chi^2$
method, which is shown as solid line in Figure~\ref{SFR}. The right
panel shows the optical depth $\tau_e$ due to the scattering between
the ionized gas and the CMB photons. The WMAP nine-year data gives
$\tau_e=0.089\pm 0.014$ \cite{Hinshaw12}, which is shown as the
shaded region. So our GRB-inferred SFR can reproduce the CMB optical
depth.

\section{Pop III GRBs and metal enrichment}

\begin{figure*}
\centering \subfigure[]{
\begin{minipage}[b]{0.8\textwidth}
\includegraphics[width=0.5\textwidth]{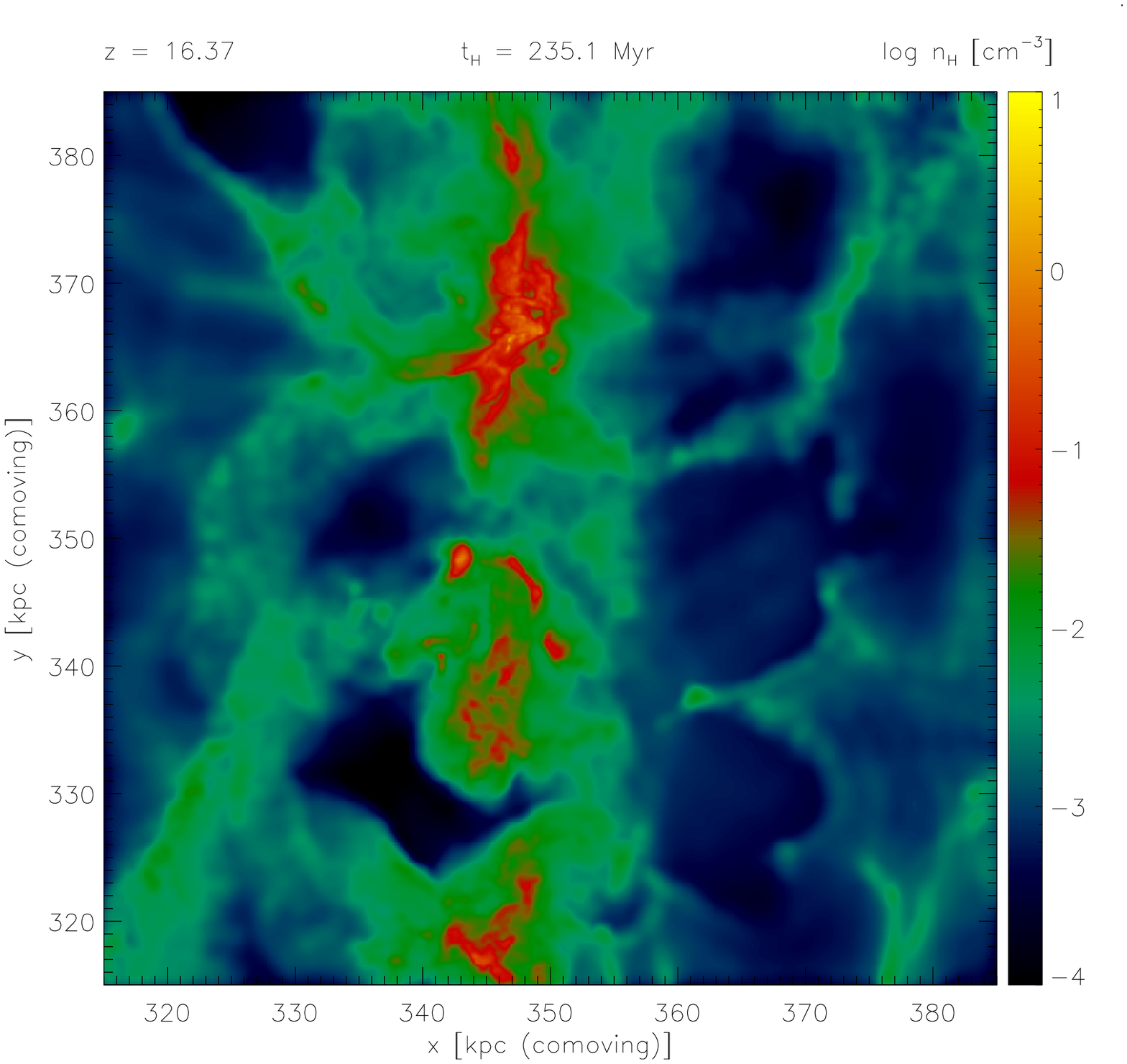}
\includegraphics[width=0.5\textwidth]{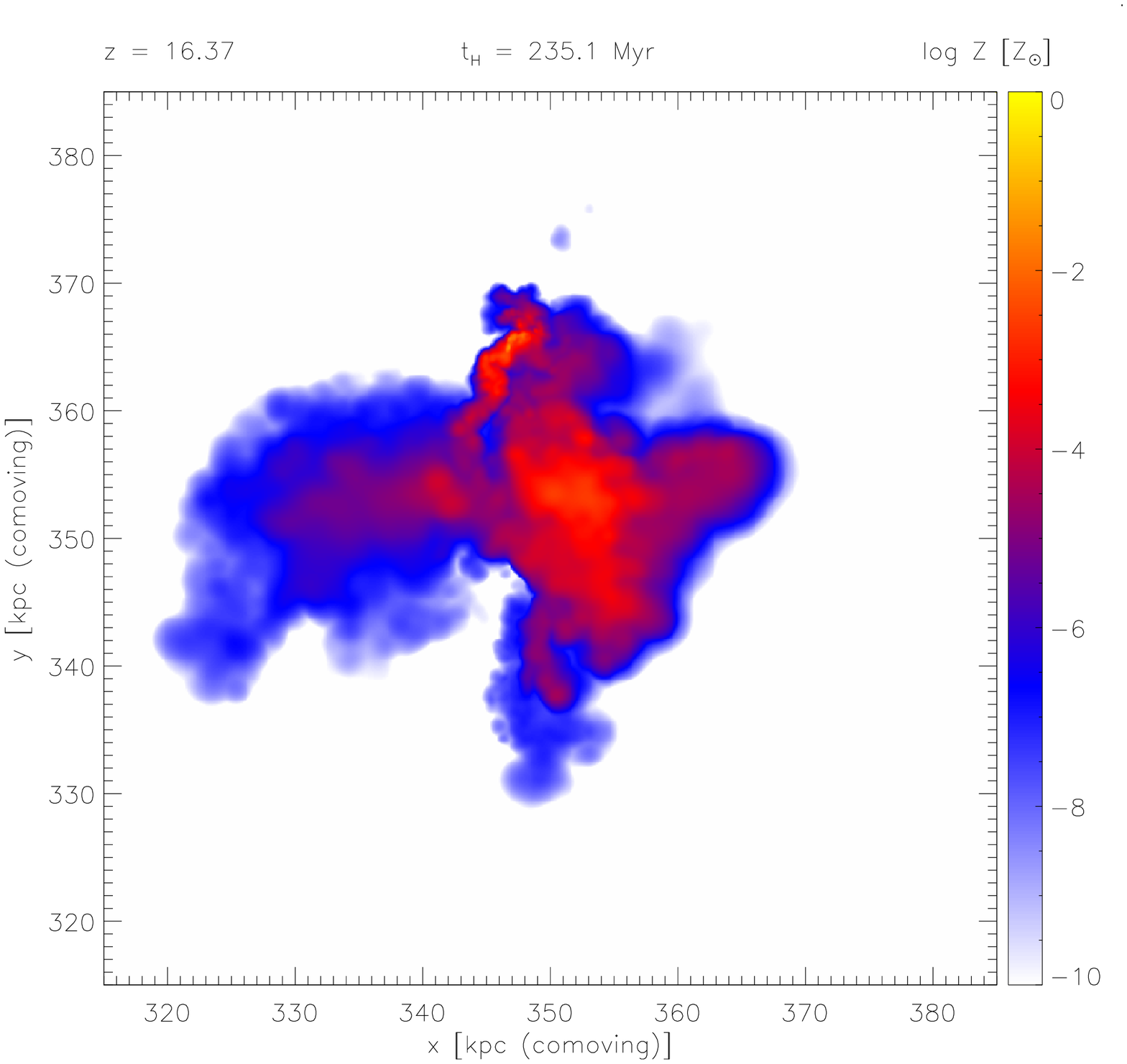}
\end{minipage}
}\caption{Possible explosion sites for Pop~III bursts. Shown are the
hydrogen number density and metallicity contours during the assembly
of a first galaxy, averaged along the line of sight within the
central $\simeq 100$\,kpc (comoving), at $z\simeq 16.4$, closer to
the virilization of the atomic cooling halo. Now, metals are being
re-assembled into the growing potential well of the first galaxy.
The topology of metal enrichment is highly inhomogeneous, with
pockets of highly enriched material embedded in regions with a
largely primordial composition. Adopted from \cite{Wang12}. }
\label{galsim}
\end{figure*}

\begin{figure}
\centering
\includegraphics[width=0.5\textwidth]{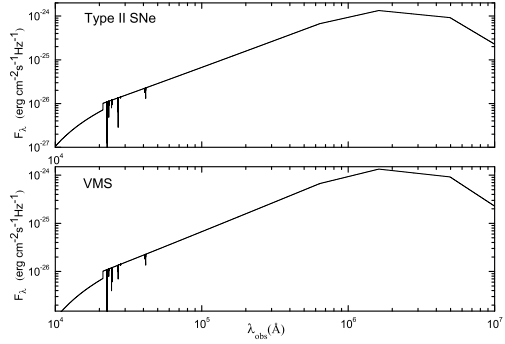} \caption{The total GRB spectrum with the
metal absorption lines at the reverse shock crossing time
($t_{\oplus}=16.7\times(1+16.37)$s). Adopted from
\cite{Wang12}.}\label{LOSItotalspeTC}
\end{figure}

Absorption lines on the spectra of bright background sources, such
as, are main sources of information about the chemical properties of
high-redshift universe \cite{Fan06}.  We simulate the formation of
first galaxy. When a GRB explodes, its light will go through the
metal-polluted region, so absorption line will appear in the
spectrum. We found that the metals in the first galaxies produced by
the first supernova explosion are likely to reside in low ionization
states (C II, O I, Si II and Fe II). The GRB afterglow travel
through the polluted environment in the first galaxy. The metals
will produce the metal absorption lines in the GRB spectrum, as
shown in Figure \ref{LOSItotalspeTC}. Because the topology of metal
enrichment could be highly inhomogeneous, so along different lines
of sight, the metal absorption lines may show a distinct signature.
The metal absorption lines, in terms of their flux density and
corresponding equivalent widths, will be detectable in the near
future, with the launch of the JWST. From Figure 4, we can
distinguish whether the Pop III progenitor dies as PISN, or Type II
supernova from the observation of Pop III GRBs spectrum
\cite{Wang12}.

\section{Self-organized criticality in GRBs}

\begin{figure}
\includegraphics[width=0.5\textwidth]{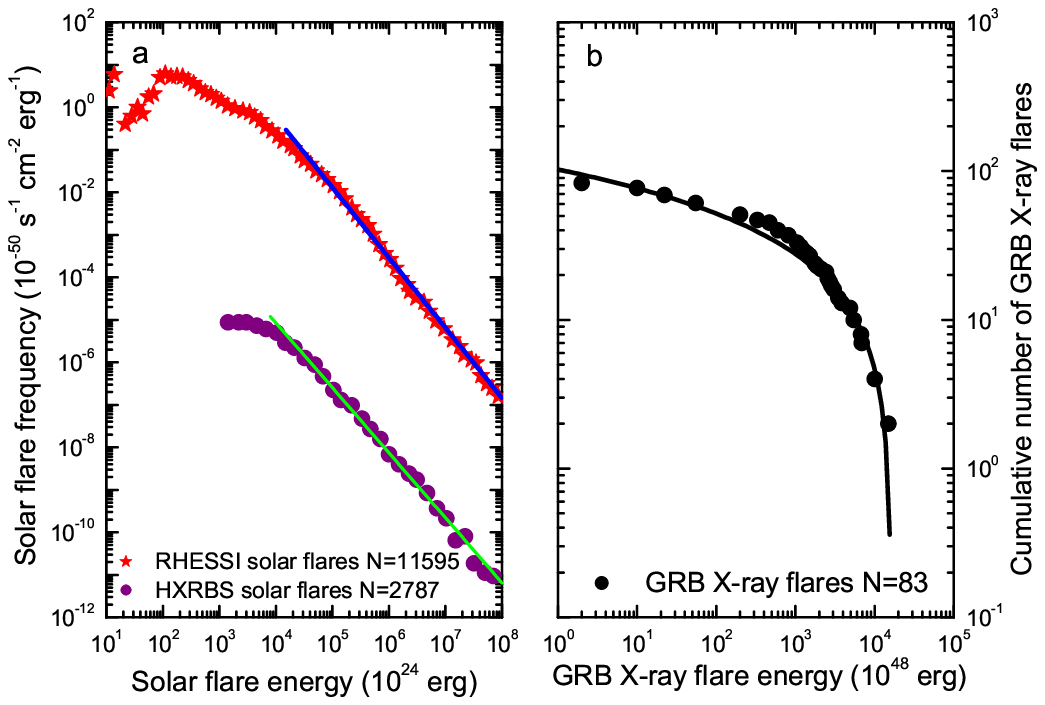}
\includegraphics[width=0.5\textwidth]{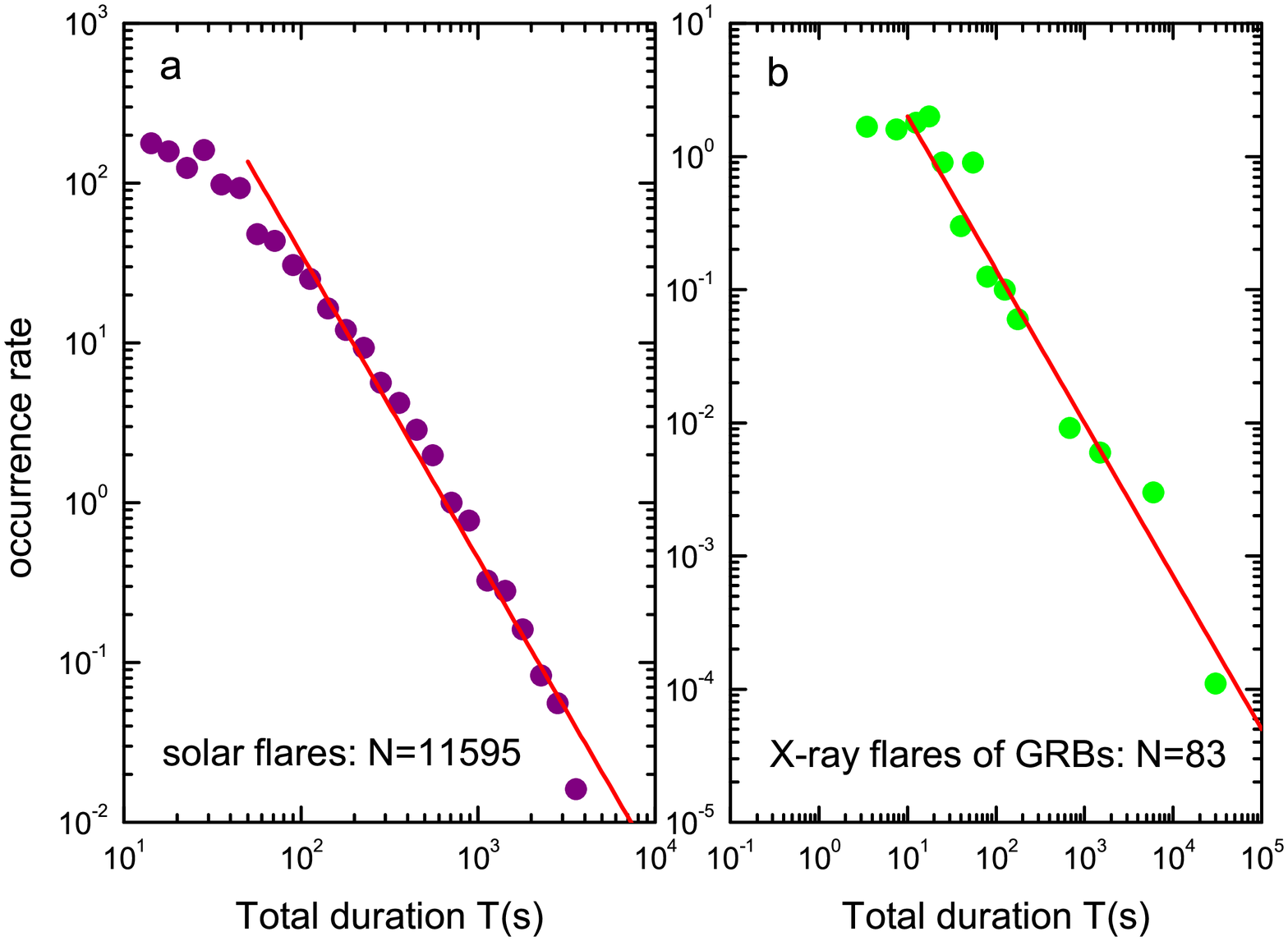}
\caption{\label{energy} \textbf{Left: Energy-release frequency
distributions.} \textbf{a}, The differential energy frequency
distribution of solar hard X-ray flares. \textbf{b}, The cumulative
energy distribution of GRB X-ray flares. The 83 GRB X-ray flares are
shown as black dots. The black curve gives the cumulative energy
distribution $N(>E)=a+b[E^{1-\alpha_E}-E_{\rm max}^{1-\alpha_E}]$
with $\alpha_E\sim 1.06\pm 0.15$ for GRB X-ray flares.
\textbf{Right: Duration-time frequency distributions}. \textbf{a},
The relation between the occurrence rate and duration time for solar
flares. \textbf{b}, The relation between the occurrence rate and
duration time for GRB X-ray flares shown as green dots. Adopted from
\cite{Wang13}.}
\end{figure}
Thanks to the rapid-response capability and high sensitivity of the
Swift satellite, numerous unforeseen features have been discovered,
one of which is that about half of bursts have large, late-time
X-ray flares with short rise and decay times. The unexpected X-ray
flares with an isotropic-equivalent energy from $10^{48}$ to
$10^{52}$ ergs have been detected for both long and short bursts. We
find that X-ray flares and solar flares share three statistical
properties: power-law frequency distributions for energies (left
panel of Figure 5), durations (right panel of Figure 5), and waiting
times, which can be explained by self-organized criticality (SOC)
theory \cite{Wang13}. Guidorzi et al. (2015) found that the waiting
time distributions of gamma-ray pulses and X-ray flares of GRBs also
had power-law tail extending over four decades, which can be
understood by SOC theory \cite{Guidorzi15}. For a SOC system
\cite{Katz86,Bak87,Bak88}, owing to some driving force, subsystems
will self-organize to a critical state at which a small perturbation
can trigger an avalanche-like chain reaction of any size within the
system. The two types of flares are driven by a magnetic
reconnection process.

\section{Conclusions}
GRBs are promising tools for studying the properties of early
Universe, including dark energy, star formation rate and metal
enrichment history. In the future, the French-Chinese satellite
Space-based multi-band astronomical Variable Objects Monitor (SVOM)
and JWST, have been optimized to increase the number of GRB and the
synergy with the ground-based facilities. This will open a new
window of the dark ages of the Universe.

\end{document}